\documentclass[12pt]{article}
\begin{document}
\title{A model of the two-dimensional quantum harmonic oscillator in an $AdS_3$ background}

\date{}
\author{Rudolf Frick\thanks{Email: rf@thp.uni-koeln.de}\\ Institut f\"ur Theoretische Physik, Universit\"at zu K\"oln,\\ Z\"ulpicher Str. 77, 50937 K\"oln, Germany }  
\maketitle

\begin{abstract}
In this paper we study a  model of the two-dimensional quantum harmonic oscillator in a 3-dimensional anti-de Sitter background. We use a generalized Schr\"odinger picture in which the analogs of the Schr\"odinger operators of the particle are independent of both the time and the space coordinates in different representations. The spacetime independent operators  of the particle induce the Lie algebra of Killing vector fields of the $AdS_3$ spacetime. In this picture, we have a metamorphosis of the  Heisenberg's uncertainty relations. 
\end{abstract}

\section{Introduction}
 In \cite{Shap} it was proposed to classify the states of a relativistic particle by means of the invariant operators (${\bf p}$ = momentum, $p_0=\sqrt{m^2c^4+c^2{\bf p}^2}$, m=mass,)
\begin{equation}
\label{1.1}
C_1({\bf p})={\bf N}^2-{\bf L}^2,{\quad}C_2({\bf p})={\bf N}\cdot{\bf L} 
\end{equation}
characterizing the infinite-dimensional unitary representations  of the Lorentz group,  and to carry out the expansion of the wave function in the momentum space representation over the functions ($0\leq\alpha<\infty$, ${\bf n}^{2}({\theta},{\varphi})=1$)
\begin{equation}
\label{1.2}
{\xi}^{(0)} ({\bf p},{\alpha},{\bf n}):=[(p_{0}-c{\bf p}\cdot{\bf n})/mc^2]^{-1+i\alpha}.
\end{equation}
The functions ${\xi}^{(0)} ({\bf p},{\alpha},{\bf n})$ are the eigenfunctions of the operator  $C_1({\bf p})$, ($C_1({\bf p}){\Longrightarrow}{1+{\alpha}}^2$).
 The boost and rotation generators of the Lorentz group have the form ($spin=0$) 
\begin{equation}
\label{1.3}
N_{i}=i{p_0}\frac{\partial}{\partial{cp_i}},{\quad} L_{i}=i{\epsilon}_{ijk}{p}_{k}\frac{\partial}{\partial{p_j}}.
\end{equation}
 The operator $C_{2}({\bf p})$ vanishes for a spinless particle.

The expansion proposed in \cite{Shap} does not include any dependence on the time $t$ and space coordinates ${\bf x}$, i.e. it is ''spacetime independent''. In \cite{Kad}, in the framework of a two-particle equation of the quasipotential type, the expansion over the functions $\xi^{\ast}({\bf p},{\alpha},{\bf n})$ was used to introduce the ``relativistic configurational'' representation (in following $\rho{\bf n}$-representation, $\rho=\alpha\hbar/mc$) . In this approach the variable $\rho$ was interpeted as the relativistic generalization of a relative coordinate. It was shown that the corresponding operators of the Hamiltonian $H(\rho,{\bf n})$ and the 3-momentum ${\bf P}(\rho,{\bf n})$, defined on the functions $\xi^{\ast}({\bf p},{\rho},{\bf n})$, has a form of the differential-difference operators.
 
In our previous papers \cite{Fri1,Fri2} it has been shown that the $\rho{\bf n}$-representation   may  also be used in a so-called generalized Schr\"odinger picture in which the analogs of the Schr\"odinger operators of a particle are independent of both the time and the space coordinates in different representations. It was found that the operators $H(\rho,{\bf n})$,  ${\bf P}(\rho,{\bf n})$,  ${\bf L}({\bf n})$ and ${\bf N}(\rho,{\bf n})={\rho}{\bf n}+({\bf n}\times{\bf L}-{\bf L}\times{\bf n})/2mc$ satisfy the commutations relations of the Poincar\'e algebra in the $\rho{\bf n}$-representation. We have two spacetime independent representation of the Poincar\'e algebra; the ${\bf p}$ and the $\rho{\bf n}$-representation. In the GS-picture  the $\rho{\bf n}$-representation  may be used to describe extendent objects like strings.

 In the case of the one-dimensional momentum space representation  $(p$ = momentum, $m$ = mass, ${p^2_0-c^2p^2}=m^2c^4)$  the eigenfunctions of the boost generator $N(p)=ip_0{\partial}_{cp}$, ($N{\Longrightarrow}{\frac{mc}{\hbar}\rho}$)  may be written in the form
\begin{equation}
\label{1.4}
{\xi}_1(p,{\rho})=[(p_0-cp)/mc^2]^{i{\frac{mc}{\hbar}\rho}}.
\end{equation}
The following expansion    
\begin{equation}
\label{1.5}
\psi({\rho})=\frac{1}{(2\pi)^{1/2}}\int\frac{d{p}}{p_0}\,\psi({p})\,{\xi^{\ast}}_1({p},{\rho}),
\end{equation}
leads to the functions $\psi({\rho})$ in the $\rho$-representation. In the $\rho$-representation the Hamilton operator $H$ and the momentum operator $P $ of the particle have the form (${\tilde\lambda}=\frac{\hbar}{mc}$)
\begin{equation}
\label{1.6}
H({\rho})=mc^2\cosh(-i{\tilde\lambda}{\partial}_{{\rho}}),{\quad}P({\rho})=mc\sinh(-i{\tilde\lambda}{\partial}_{{\rho}}),
\end{equation}
and satisfy the commutation relations of the Poincar\'e algebra
\begin{equation}
\label{1.7}
\lbrack{\rho},P\rbrack=i\frac{\hbar}{mc^2}H,\quad\lbrack{P},{H}\rbrack=0,\quad\lbrack{H},{\rho}\rbrack=-i\frac{\hbar}{m}P.
\end{equation}
For a free particle in the Minkowski spacetime of two dimensions (d=2), the coordinates $t$, ${x}$ may be  introduced in the states   with the help of the transformation  
\begin{equation}
\label{1.8}
 S(t,{x})=\exp[-i(tH -xP)/\hbar].
\end{equation}  
 We obtain
\begin{equation}
\label{1.9}
S(t,{x})\psi({\rho})=\psi({\rho},t,{x})=\frac{1}{(2\pi)^{1/2}}
\int\frac{d{p}}{p_0}\,\psi({p},t,{x})\,{\xi^{\ast}_1}({p},{\rho}),
\end{equation}
where $\psi(p,t,x)=\psi(p)exp[-i(tp_0-xp)/\hbar]$. 

In the case of a point particle ($\rho=0$)  we have the Fourier transform in  relativistic quantum mechanics
\begin{equation}
\label{1.10}
\psi(t,x)=\frac{1}{(2\pi)^{1/2}}\int\frac{d{p}}{p_0}\,\psi(p,t)e^{\frac{ixp}{\hbar}}.
\end{equation}

In (\ref{1.9}),  the  spacetime coordinates appear in the states in  the $\rho$ and in the ${p}$-representation. We have a metamorphosis of the  Heisenberg's uncertainty relation ${\Delta}x{\cdot}{\Delta}p\geq{\hbar}/2$. From $\lbrack{\rho},P\rbrack=i{\hbar}H/mc^2$ in (\ref{1.7}) follows that instead of ${\Delta}x{\cdot}{\Delta}p\geq{\hbar}/2$, we have
\begin{equation}
\label{1.11}
{\Delta\rho\cdot\Delta{p}\geq\hbar}/2.
\end{equation}

 The GS-picture may be used in a quantum theory of gravity in which objects need a sharply defined frame. In the paper \cite{Fri2},  this picture was used  to describe the motion of  a relativistic particle in anti-de Sitter spacetime (d=2, d=4). It  was found that the spacetime independent operators of the particle in an external field (like in the case of a harmonic oscillator) induce the Lie algebra of Killing vector fields of the $AdS_4$ spacetime $(d=4;a=1,2,...10;$ $\lbrace{x^{i}\rbrace}$, $i=1,2,3.)$
\begin{equation}
\label{1.12}
{K}_a(t,x^{i})\widetilde{\Phi}({\rho},{\bf n},t,x^{i})=B _a({\rho},{\bf n})\widetilde{\Phi}({\rho},{\bf n},t,x^{i}).
\end{equation}
Here $\widetilde{\Phi}$ denotes the wave function of the particle. The operators of Killing vector field  ${K}_a(t,x^{i})$ satisfy  the same commutation rules as  the spacetime independent operators $B _a({\rho},{\bf n})$, except for the minus signs on the right-hand sides. The equations (\ref{1.12}) are valid for any d. In the present paper we use these equations to describe the motion of a particle in $AdS_3$ spacetime. In the case of d = 3 we need  six spacetime independent operators of the particle.  In Sect. 2  we will now show that the operators of  a relativistic  model of the two-dimensional quantum harmonic oscillator in the ${\rho}{\bf n}$-representation can be used in  Eqs. (\ref{1.12}).  This  will allow us to obtain an exact expression for the energy  levels of the particle and an expression for  the spectrum of the $AdS_3$ radius.

\section{One particle quantum equation in $AdS_3$ \\spacetime} 

In the two-dimensional  momentum  space representation, the first Casimir operator of the Lorentz group $C_1({\bf p})$ has the eigenfuctions
($p^2_0-c^2p^2_1-c^2p^2_2=m^2c^4$)
\begin{equation}
\label{1.13}
\xi_{2}({\bf p},{\rho},{\bf n}):=[(p_{0}-c{\bf p}\cdot{\bf n})/mc^2]^{-\frac{1}{2}+i\frac{mc}{\hbar}\rho},
\end{equation}
where $(n_1={\cos\varphi}, n_2={\sin\varphi})$.

The Hamilton operator  and the momentum operators of the particle defined on the functions ${\xi^{*}}_2(\bf p,\rho,\bf n)$ have the form \cite{Nag}
\begin{equation}
\label{1.14}
H(\rho,{\bf n})=mc^2\cosh\left(i{\tilde\lambda}{\partial}_{\rho}\right)+\frac{i\hbar{c}}{2\rho}\sinh\left(i{\tilde\lambda}{\partial}_{\rho}\right) -\frac{({-i{\hbar}{\partial}_{\varphi}})^2}{m\rho(2\rho+i\tilde\lambda)}e^{i\tilde\lambda{\partial}_{\rho}},
\end{equation}
 \begin{equation}
\label{1.15}
{P_1}(\rho,{\bf n})=mc{n_1}(H/mc^2-e^{i\tilde\lambda{\partial}_{\rho}})+\frac{i\hbar{n_2}\cdot{\partial}_{\varphi}}{\rho+2i{\tilde\lambda}}e^{i\tilde\lambda{\partial}_{\rho}},
\end{equation}
\begin{equation}
\label{1.16}
{P_2}(\rho,{\bf n})=mc{n_2}(H/mc^2-e^{i\tilde\lambda{\partial}_{\rho}})-\frac{i\hbar{n_1}\cdot{\partial}_{\varphi}}{\rho+2i{\tilde\lambda}}e^{i\tilde\lambda{\partial}_{\rho}}.
\end{equation}
The operators $H$, ${\bf P}$,  and the three operators of the Lorentz algebra in the $\rho{\bf n}$-representation
\begin{equation}
\label{1.17}
{N_1}(\rho,{\bf n})=n_1({\rho}-\frac{i}{2}\tilde\lambda)-i\tilde\lambda{n_2}{\partial}_{\varphi},
\end{equation}
\begin{equation}
\label{1.18}
{N_2}(\rho,{\bf n})=n_2({\rho}-\frac{i}{2}\tilde\lambda)+i\tilde\lambda{n_1}{\partial}_{\varphi},\quad{L}=-i{\hbar}{\partial}_{\varphi},
\end{equation}
satisfy the commutation  relations of the Poincare algebra.

For the particle in an  external field like the two-dimensional harmonic oscillator potential we use  the following operators  
\begin{equation}
\label{1.19}
\hat{P_0}(\rho,{\bf n})=H(\rho,{\bf n})+H^{'}_0(\rho),
\end{equation}
\begin{equation}
\label{1.20}
\hat{P}_i(\rho,{\bf n})=P_i(\rho,{\bf n})+P^{'}_i(\rho,{\bf n}),
\end{equation}
where $(\omega = frequency, i=1,2)$
\begin{equation}
\label{1.21}
H^{'}_0=\frac{m{\omega}^2}{2}\left(\rho-\frac{i}{2}\tilde\lambda\right)\left(\rho-i\tilde\lambda\right)e^{-i\tilde\lambda{\partial}_{\rho}},
\end{equation}
\begin{equation}
\label{1.22}
P^{'}_i= n_i\frac{m{\omega}^2}{2c}\left(\rho-\frac{i}{2}\tilde\lambda\right)\left(\rho-i\tilde\lambda\right)e^{-i\tilde\lambda{\partial}_{\rho}}.
\end{equation}

In the nonrelativistic limit the operator $\hat{P_0}(\rho)-mc^2$ assume the form
\begin{equation}
\label{1.23}
\hat{P_0}_{\rm{nr}}=-\frac{{\hbar}^2}{2m}\frac{{\partial}^2}{\partial{\rho}^2}+\frac{{{\hbar}^2{\partial}_{\varphi}}^2}{2m{\rho}^2}+\frac{m{\omega}^2}{2}{\rho}^2.
\end{equation}
The operators 
\begin{equation}
\label{1.24}
\hat{P_0}(\rho,{\bf n})=H(\rho,{\bf n})+H^{'}_0(\rho,{\bf n}),
\end{equation}
\begin{equation}
\label{1.25}
\hat{P}_i(\rho,{\bf n})=P_i(\rho,{\bf n})+P^{'}_i(\rho,{\bf n}),
\end{equation}
 and ${L}$, ${N}_i(\rho,{\bf n})$ satisfy the  commutations rules of the Lie algebra $so(2,2)$
\begin{equation}
\label{1.26}
\lbrack{N_i},\hat{P}_j\rbrack=\frac{\imath\hbar}{mc^2}\delta_{ij}\hat{P_0},\quad\lbrack\hat{P}_i,\hat{P_0}\rbrack=-\imath\hbar{m}{\omega}^2{N_i}\quad\lbrack\hat{P_0},{N_i}\rbrack=-\frac{\imath\hbar}{m}\hat{P_i},
\end{equation}
\begin{equation}
\label{1.27}
\lbrack\hat{P}_1,\hat{P}_2\rbrack=-\imath\hbar\frac{{\omega}^2}{c^2}{L},\quad\lbrack{L},\hat{P_0}\rbrack=0,\quad\lbrack\hat{P}_1,L\rbrack=-\imath\hbar\hat{P}_2,
\end{equation}
\begin{equation}
\label{1.28}
\lbrack{N_1},{N_2}\rbrack=-\frac{\imath\hbar}{m^2c^2}{L},\quad\lbrack{N_1},{L}\rbrack=-\imath\hbar{N_2}.
\end{equation}
For the  Casimir operator 
 \begin{equation}
\label{1.29}
C(\rho,{\bf n})=\frac{1}{(\hbar\omega)^2}\left\lbrace\hat{P_0}^{2}-c^2{\bf P}^{2} \right\rbrace-\frac{m^2c^2}{{\hbar}^2}{\bf N}^2+\frac{1}{{\hbar}^2}{L}^2,
\end{equation} 
we have
\begin{equation}
\label{1.30}
C(\rho,{\bf n})=(\frac{m^2c^4}{{\hbar}^2{\omega}^2}-3/4){I}.
\end{equation} 

The explicit forms of the six operators ${K}_a(t,x^{i})$ depend on the realisation in terms of the spacetime coordinates. We have the problem of determining  observables in the GS-picture. In order to interpret the operator $\hat{P_0}$ as  Hamilton operator, we chose the following realisation ($t$, $x_1$, $x_2$ ($x_1=r\cos\widetilde{\varphi}$,  $x_2=r\sin\widetilde{\varphi}$,  $i=1,2$)),
\begin{equation}
\label{1.31}
K_{03}=i\hbar\frac{\partial}{\partial{t}},
\end{equation}
\begin{equation}
\label{1.32}
K_{i3}=\sqrt{1+(\omega{r}/c)^2}\cos\omega{t}({i\hbar\partial}_{x_i})-\frac{(\omega{x_i/c^2})\sin\omega{t}}{\sqrt{1+(\omega{r}/c)^2}}i{\hbar\partial}_{t}.
\end{equation}
\begin{equation}
\label{1.33}
K_{i0}=\frac{1}{m\omega}\sqrt{1+(\omega{r/c)^2}}\sin\omega{t}({i\hbar\partial}_{x_i})+\frac{x_i\cos\omega{t}}{mc^2\sqrt{1+(\omega{r/c})^2}}i{\hbar\partial}_{t},
\end{equation}
\begin{equation}
\label{1.34}
K_{12}=i\hbar\left(x_1\frac{\partial}{\partial{x_2}}-x_2\frac{\partial}{\partial{x_1}}\right)=i{\hbar\partial}_{\widetilde{\varphi}}
\end{equation}
The set of the operators $\lbrace{K_{03},K_{i3},K_{i0},K_{ij}}\rbrace$ determines the same Lie algebra  as the operators $\lbrace\hat{P_0},\hat{P}_i,N_i,L\rbrace$ except for the minus signs on the right-hand sides
\begin{equation}
\label{1.35}
\lbrack{K_{i0},K_{j3}}\rbrack=-\frac{\imath\hbar}{mc^2}\delta_{ij}K_{03},\quad\lbrack{K_{i3},K_{03}}\rbrack=\imath\hbar{m}{\omega}^2{K_{i0}},\quad\lbrack{K_{03},K_{i0}}\rbrack=\frac{\imath\hbar}{m}{K_{i3}},
\end{equation}
\begin{equation}
\label{1.36}
\lbrack{K}_{i3},{K}_{j3}\rbrack=\imath\hbar\frac{{\omega}^2}{c^2}K_{ij},\quad\lbrack{K_{ij},K_{03}}\rbrack=0,\quad\lbrack{K}_{i3},K_{ik}\rbrack=\imath\hbar{K}_{k3},
\end{equation}
\begin{equation}
\label{1.37}
\lbrack{K_{i0},K_{j0}}\rbrack=\frac{\imath\hbar}{m^2c^2}K_{ij},\quad\lbrack{K_{i0},K_{ik}}\rbrack=\imath\hbar{K_{k0}}.
\end{equation}
The operators $\lbrace{K_{03},K_{i3},K_{i0},K_{ij}}\rbrace$ form a basis for the $SO(2,2)$ group generators and related to  Killing vectors of the  $AdS_3$ spacetime with metric  
\begin{equation}
\label{1.38}
ds^2=\left(1+\frac{{\omega}^2r^2}{c^2}\right)c^2dt^2-\frac{1}{1+\frac{{\omega}^2r^2}{c^2}}dr^2-r^2{d^2{\widetilde{\varphi}}}.
\end{equation}
Here, the constant $\omega/c$ is related to the radius $\kappa$ of the  $AdS_3$ spacetime ($\kappa=c/{\omega}$).

We can introduce the equation 
\begin{equation}
\label{1.39}
i\hbar\frac{\partial}{\partial{t}}\Phi(\rho,{\bf n};t,x_1,x_2)=\hat{P_0}(\rho,{\bf n})\Phi(\rho,{\bf n};t,x_1,x_2),
\end{equation}
which define the operator $\hat{P_0}(\rho,{\bf n})$ as Hamilton operator of the particle.

 A general solution of $\Phi(\rho,{\bf n};t,x_1,x_2)$  can be written as a sum of separated solutions or the eigenfunctions of the operators $\hat{P_0}(\rho,{\bf n})$ and the Casimir operator $(\tau={\omega}t$, $\tan\sigma={\omega}r/c)$
\begin{equation}
\label{1.40}
C(\tau,\sigma,\widetilde{\varphi})=-\cos^2\sigma\frac{{\partial}^2}{\partial{\tau}^2}+\cot\sigma\frac{\partial}{\partial\sigma}+\cos^2\sigma\frac{{\partial}^2}{\partial{\sigma}^2}+\cot^2\sigma\frac{{\partial}^2}{\partial{{\widetilde{\varphi}}^2}}.
\end{equation}
The eigenfunctions  of $C(\tau,\sigma,\widetilde{\varphi})$   are  $(n=0,1,2.., {\vert\widetilde{m}\vert}=0,1,2.., \\ M=2,3,4,..,  \lambda=2n+{\vert\widetilde{m}\vert}+M)$
\begin{equation}
\label{1.41}
{\psi}^{M}_{n{\vert\widetilde{m}\vert}\widetilde{m}}=N^{M}_{n{\vert\widetilde{m}\vert}\widetilde{m}}e^{-i\lambda\tau}(\cos\sigma)^{M}(\sin\sigma)^{\vert\widetilde{m}\vert}{_2}F_1(a,b,c;z))e^{i\widetilde{m}\widetilde{\varphi}},
\end{equation}
where $N^{M}_{n{\vert\widetilde{m}\vert\widetilde{m}}}$ are the normalization constants and 
\begin{equation}
\label{1.42}
{_2}F_{1}(\frac{1}{2}({\vert\widetilde{m}\vert}+M-\omega),\frac{1}{2}({\vert\widetilde{m}\vert}+M+\omega),3/2+{\vert\widetilde{m}\vert};{\sin^2\sigma})
\end{equation}
 are the hypergeometric functions.
Thus we find
\begin{equation}
\label{1.43}
C(\tau,\sigma,\widetilde{\varphi}){\Rightarrow}M(M-2).
\end{equation}
For the spectrum of $i\hbar\frac{\partial}{\partial{t}}$ we have
\begin{equation}
\label{1.44}
E=\hbar\omega(2n+{\vert\widetilde{m}\vert}+M). 
\end{equation}

The eigensolutions  of the Hamilton  operator $\hat{P_0}(\rho,{\bf n})$
\begin{equation}
\label{1.45}
\hat{P_0}(\rho,{\bf n}){\xi}=\hbar\omega(2n+{\vert\widetilde{m}\vert}+M){\xi}
\end{equation}
are $(\widetilde\rho=\frac{mc\rho}{\hbar}, M=1+\sqrt{1/4+(\frac{mc^2}{\hbar\omega}})^2, {l}= \vert{m}\vert=0,1,2..,)$
\begin{equation}
\label{1.46}
{\xi}^{M}_{nlm}=c^{M}_{nl}(\frac{\hbar\omega}{mc^2})^{-i\widetilde\rho}\Gamma(M-1/2-i\widetilde\rho)\frac{\Gamma[l+1/2+i\widetilde\rho)}{\widetilde\Gamma(i\widetilde\rho+1/2)}S_{n}(\widetilde\rho){e^{im\varphi}},
\end{equation}
where  $c^{M}_{nl}$ are normalization constants and $S_{n}(\widetilde\rho)$ are the Hahn polynomials
\begin{eqnarray}
\label{1.47}
S_{n}({\widetilde\rho}^2;l+1/2,M-1/2),1/2))=\frac{\Gamma(l+M+n)\Gamma(l+M)}{\Gamma(l+1+n)\Gamma(l+1)}\nonumber\\{\times}{_3}F_2(-n,l+1/2+i{\widetilde\rho},l+1/2-i{\widetilde\rho};l+M,l+1;1).
\end{eqnarray}

For the function $\Phi(\rho,{\bf n};t,x_1,x_2)$, we have ($m=\widetilde{m}$, $l={\vert\widetilde{m}\vert}$)
\begin{equation}
\label{1.48}
{\Phi}^{M}=\sum_{n=0,l=0}^{\infty}{\sum_{m=-l}^{m=l}}{\psi}^{M}_{n{\vert\widetilde{m}\vert}\widetilde{m}}    {\xi}^{M}_{nlm}.
\end{equation}
From  
\begin{equation}
\label{1.49}
C(\rho,{\bf n}){\Phi}^{M}=C(\tau,\sigma,\widetilde{\varphi}){\Phi}^{M}
\end{equation}
follows that the oscillator frequency is discrete and for higher $M$ decreases accoding to
\begin{equation}
\label{1.50}
{\omega}_M=\frac{mc^2}{\hbar\sqrt{(M^2-2M+3/4))}}.
\end{equation}
The energy spectrum of the particle can be written as
\begin{equation}
\label{1.51}
E_{n{\vert\widetilde{m}\vert}M}=\frac{mc^2}{\sqrt{(M^2-2M+3/4)}}(2n+\vert\widetilde{m}\vert+M).
\end{equation}
 For the $AdS_3$ radius $\kappa=c/{\omega}$, we have ${\kappa}_M=\sqrt{(M^2-2M+3/4)}$ $\hbar/mc$.

\section{Conclusion}
In this paper we have shown that a generalized Schr\"odinger picture may be used to describe a  relativistic particle in  a 3-dimensional anti-de Sitter spacetime.  A specific feature of this picture is that the frame itself becomes dynamical. It was found that in this picture we have a metamorphosis of the  Heisenberg's uncertainty relations.  We have shown that, the energy of the particle  and the anti-de Sitter radius are discrete. 

\end{document}